\documentclass[a4paper,11pt]{article}

\usepackage{contribution}

\newcommand{\weblink}[2][]{%
    \ifthenelse{\equal{#1}{}}%
    {\textnormal{\url{#2}}}%
    {\textnormal{\href{#2}{#1}}}%
}

\newcommand{\acknowledgements}[1]{%
  \bigskip\bigskip
  \textsf{\textbf{\Large Acknowledgements}} \\[2ex]
  {#1}
  \bigskip
}

\def\beq{\begin{equation}}
\def\eeq#1{\label{#1}\end{equation}}
\def\eeqn{\end{equation}}

\def\beqa{\begin{eqnarray}}
\def\eeqa#1{\label{#1}\end{eqnarray}}
\def\eeqan{\end{eqnarray}}

\let\bar=\overbar

\def\Dslash{\not{\hbox{\kern-4pt $D$}}}
\def\dslash{\not{\hbox{\kern-2pt $\del$}}}

\def\msb{{\bar{\ssstyle M \kern -1pt S}}}

\newcommand{\contribution}[7][]{%
  \clearpage
  \thispagestyle{plain}
  \ifthenelse{\equal{#1}{}}
  {\hypersetup{pdftitle={#2}}}
  {\hypersetup{pdftitle={#1}}}
  \hypersetup{pdfauthor={{#3} {#4}}}
  {\centering\normalfont\LARGE\bfseries\sffamily #2 \par\nobreak}
  \lhead{}
  \chead{%
    \textit{\footnotesize XIV International Conference on Hadron Spectroscopy
      (\weblink[\textit{hadron2011}]{http://www.hadron2011.de}), 13-17 June 2011, Munich, Germany}%
  }
  \rhead{}
  \bigskip
  \begin{center}
    {#3} {#4}\ifthenelse{\equal{#6}{}}{}{\footnote{\weblink[#6]{mailto:#6}}}
    \ifthenelse{\equal{#7}{}}{}{#7} \\
    \textit{#5}
  \end{center}
  \bigskip
}

\renewcommand{\abstract}[1]{%
  \begin{center}
    \begin{minipage}{0.85\textwidth}
      \begin{footnotesize}
        #1
      \end{footnotesize}
    \end{minipage}
  \end{center}
  \bigskip
}

\begin{document}

%
%
%
%
%
{  


%

\contribution[Baryon Spectroscopy at COMPASS]  
{Baryon Spectroscopy at COMPASS}  
{Alex}{Austregesilo}  
{Technische Universit\"at M\"unchen, Physik Department E18, D-85748 Garching}  
{aaust@tum.de}  
{on behalf of the COMPASS Collaboration}  
%

\abstract{%
  At the COMPASS experiment, diffractive dissociation of the beam proton is one of the dominant processes for the $190\,\mathrm{GeV}/c$ positive hadron beam impinging on a liquid hydrogen target. The status of the analysis of the reactions $p\,p\rightarrow p_f\,\pi^+\pi^-\,p_s$  and $p\,p\rightarrow p_f\,K^+K^-\,p_s$ is presented, where dominant features of the light-baryon spectrum become clearly visible. Furthermore, partial-wave analysis techniques to disentangle these spectra are discussed.

}
%

\section{Introduction}

COMPASS~\cite{com07} is a fixed-target experiment at the CERN SPS for the investigation of structure and spectroscopy of hadrons. The experimental setup features a large-acceptance and high-resolution spectrometer including particle identification and calorimetry and is therefore ideal to address a broad range of different final states. The $190\,\mathrm{GeV}/c$ positive hadron beam impinging on a liquid hydrogen target gives the unique possibility to study proton diffractive dissociation. This peripheral scattering process is characterised by its four-momentum transfer distribution, the slope of which approximately reflects the size of the target particles. The prerequisite of coherent production translates into an upper limit for the mass of diffractively produced resonances of a few $\mathrm{GeV}/c^2$\cite{ama76}. Exclusive events with three charged particles in the final state have been selected, this data set is the starting point for a dedicated partial-wave analysis.


\section{Event Selection}

The presented data correspond to 30\% of the recorded proton-beam data set. The events were triggered by a coincidence between the incoming beam and the recoiling proton $p_{s(low)}$ from the reaction. The dedicated recoil-proton detector (RPD) around the target measured a pure proton signal. It can therefore be safely assumed that the target protons remain intact. On the other hand, the interaction is required to have a squared four-momentum transfer $t'$ to the recoil proton larger than 0.07$\,\mathrm{GeV}^2/c^2$ in order to fall within the acceptance of the RPD trigger.

As the positive secondary hadron beam at $190\,\mathrm{GeV}/c$ consists of a mixture of $75$\% protons, $24$\% pions, and less than $1$\% kaons, the incoming beam particles were identified by two CEDAR detectors (ChErenkov Differential counter with Achromatic Ring focus) which achieved a nearly complete separation. In addition, particle identification was applied to distinguish between the fast proton $p_f$ and the positive meson in the final state. As the COMPASS RICH (Ring Imaging CHerenkov) detector does not allow proton identification directly in a large part of the kinematic range, $\pi^+$ and $K^+$ signals were used, respectivly.

Only exclusive events were selected, where all particles in the reaction were detected and their energy as well as charge sum match the incident beam. In addition, the recoil proton and the forward going three-body system ($p_f\pi^+\pi^-$ or $p_fK^+K^-$) had to be back-to-back in the plane transverse to the beam. The resulting data sample includes merely a negligible contribution of non-exclusive background.

\mathversion{bold}
\section{Diffractive dissociation of protons into $p_f\,\pi^+\pi^-$ final states}
\mathversion{normal}
\label{sec:pion}

\begin{figure}[b]
  \begin{minipage}[]{.5\textwidth}
    \centering
    \includegraphics[clip,trim= 10 0 40 10, width=.7\textwidth]{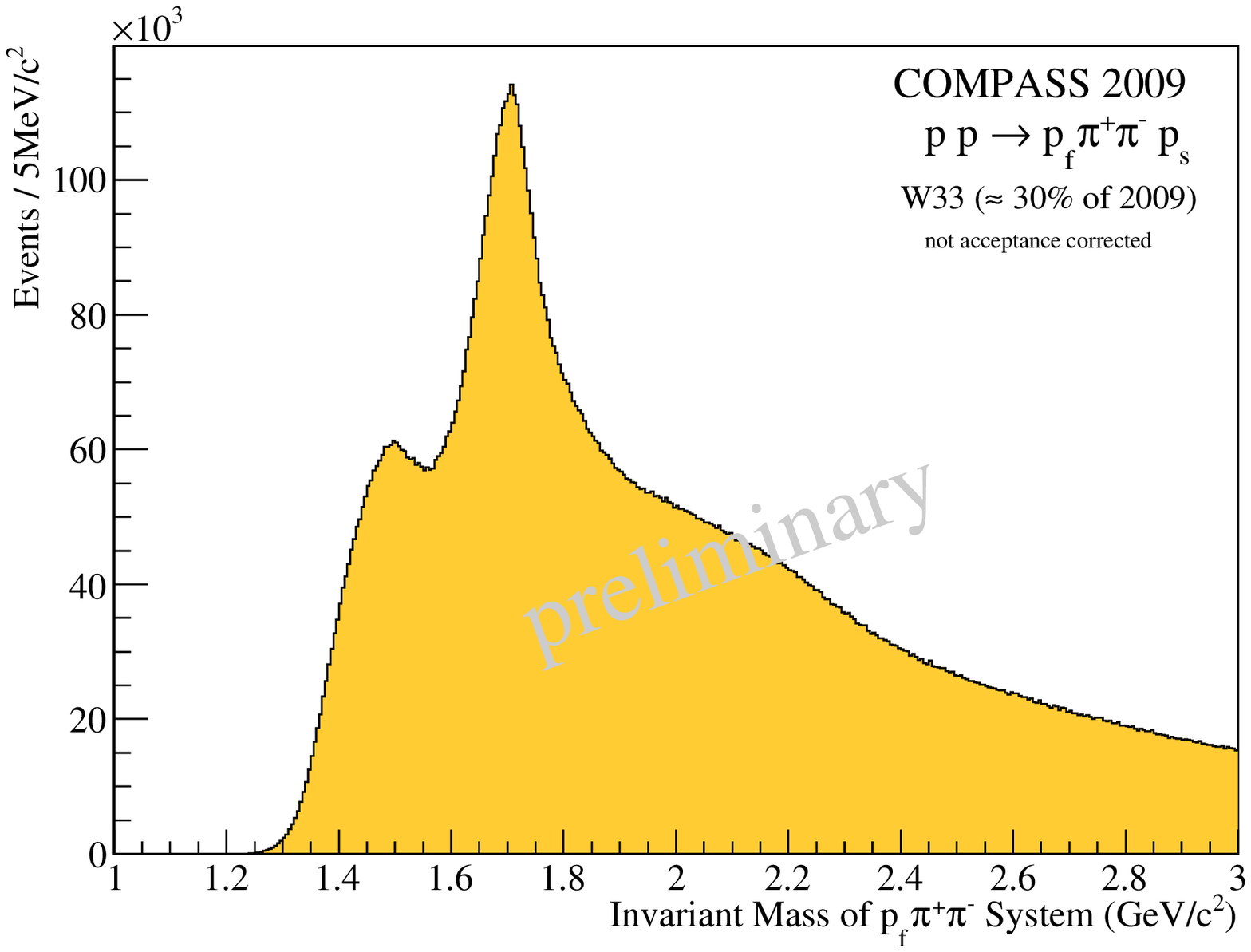}
    \caption{\em Invariant mass distribution\\ of $p_f\pi^+\pi^-$ system}
    \label{fig:m3}
  \end{minipage}
  \hfill
  \begin{minipage}[]{.5\textwidth}
    \centering
    \includegraphics[clip,trim= 10 0 40 10, width=.7\textwidth]{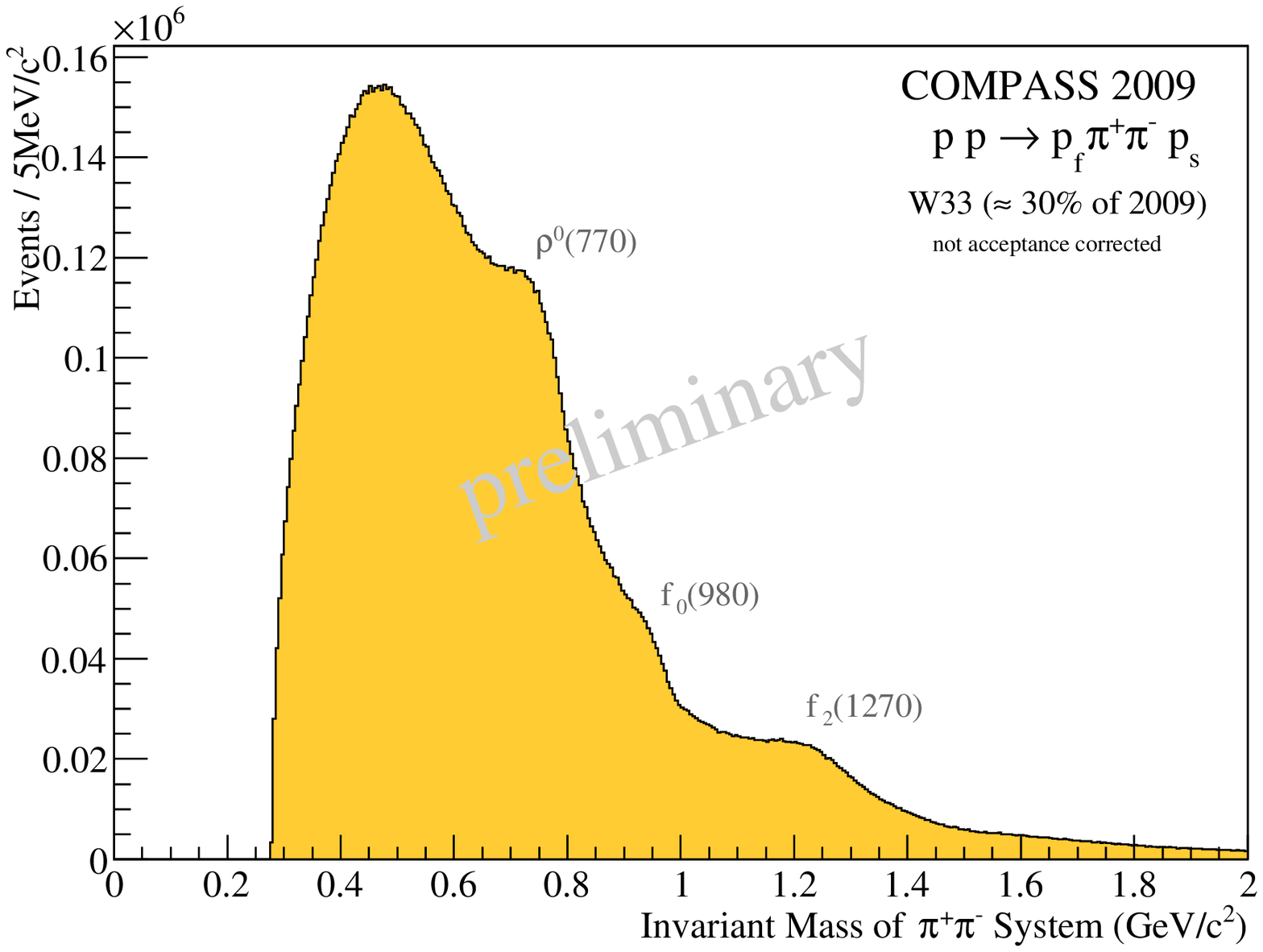}
    \caption{\em Invariant mass distribution\\ of $\pi^+\pi^-$ subsystem}
    \label{fig:mrho}
  \end{minipage}
\end{figure}

\begin{figure}
  \begin{minipage}[]{.5\textwidth}
    \centering
    \includegraphics[clip,trim= 10 0 40 10, width=.7\textwidth]{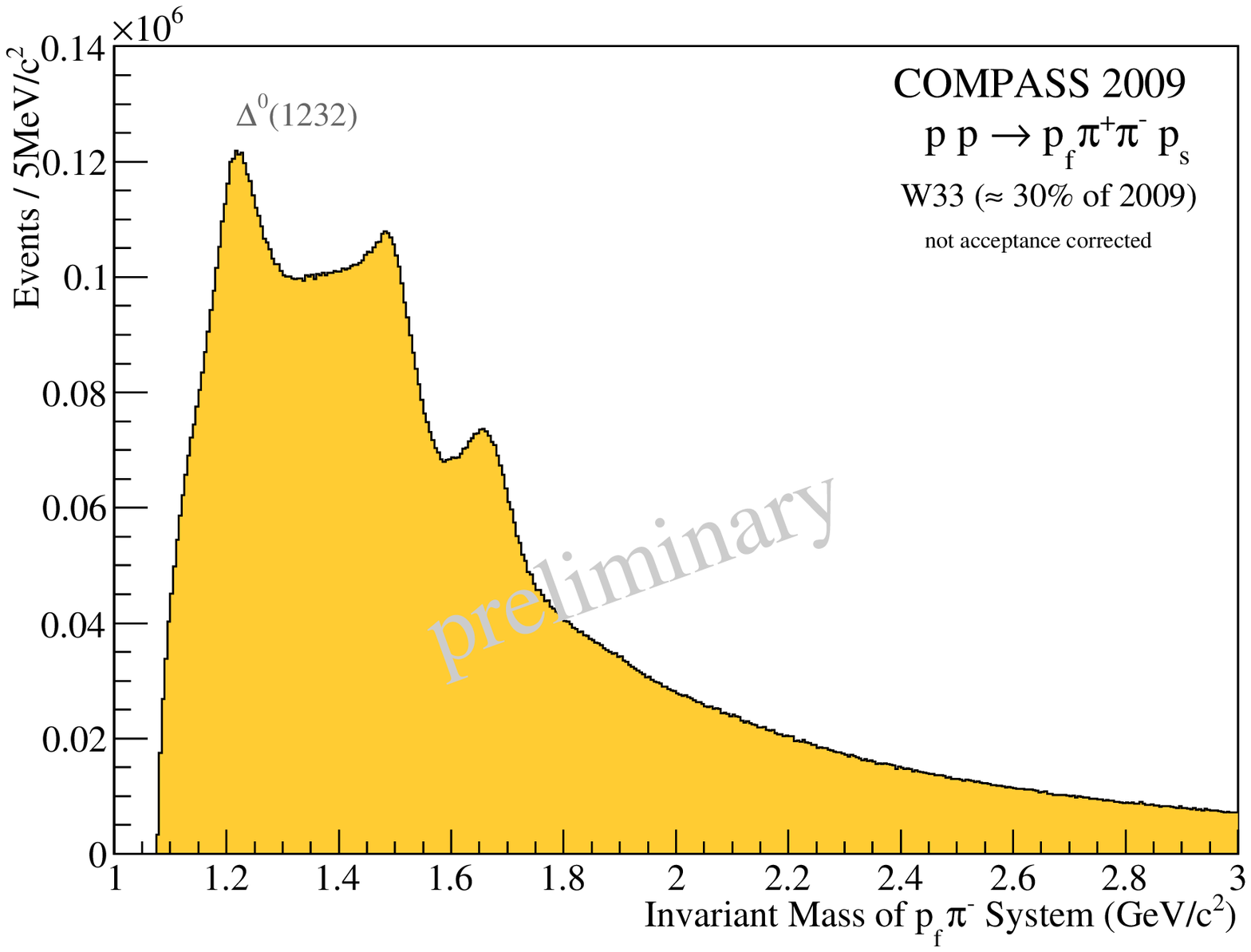}
    \caption{\em Invariant mass distribution\\ of $p_f\pi^-$ subsystem}
    \label{fig:mdeltan}
  \end{minipage}
  \hfill
  \begin{minipage}[]{.5\textwidth}
    \centering
    \includegraphics[clip,trim= 10 0 40 10, width=.7\textwidth]{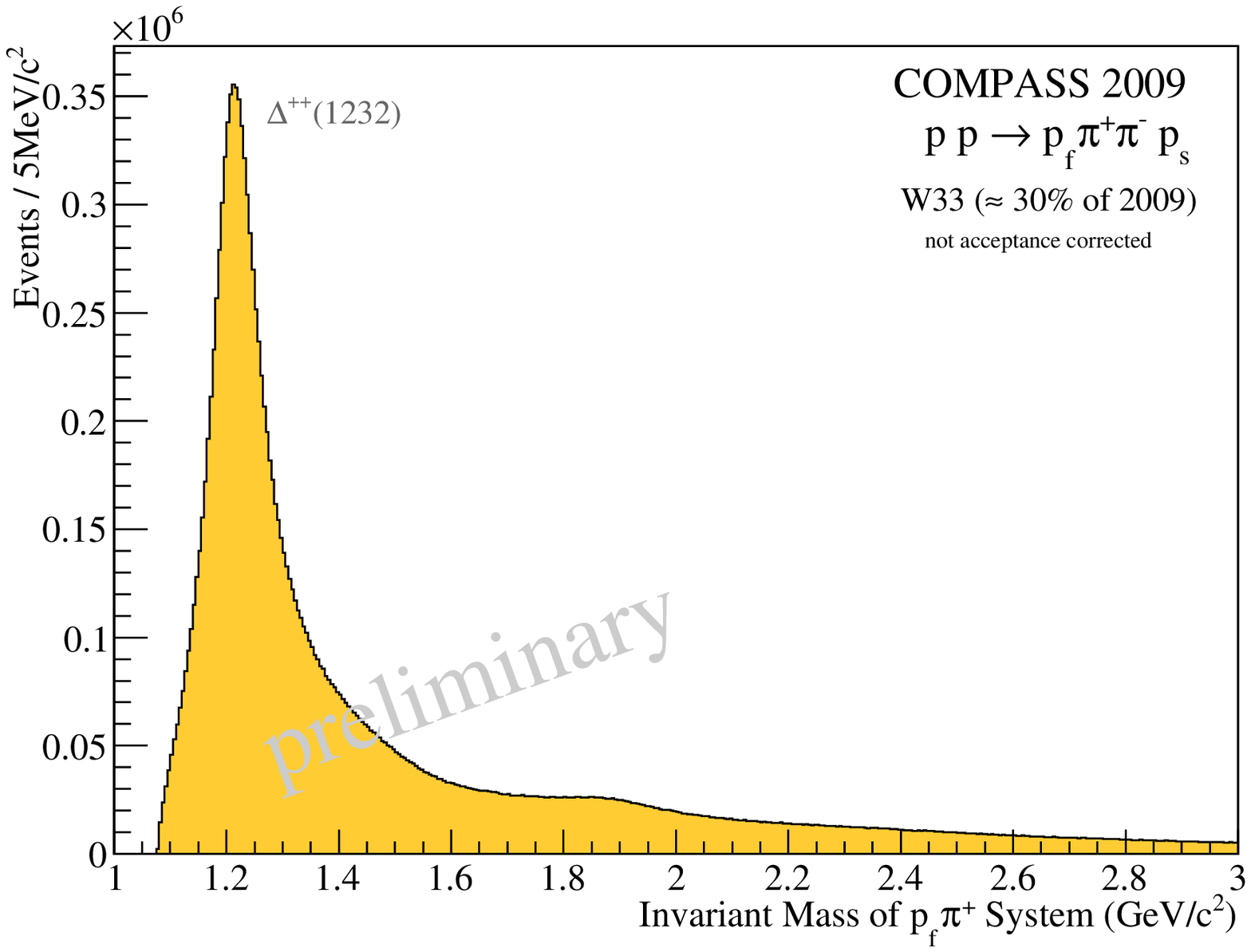}
    \caption{\em Invariant mass distribution\\ of $p_f\pi^+$ subsystem}
    \label{fig:mdeltapp}
  \end{minipage}
\end{figure}

In Fig.~\ref{fig:m3} the invariant mass distribution of the $p_f\pi^+\pi^-$ system is shown. This excited proton spectrum is foreseen to be studied in detail by the means of partial-wave analysis. Few distinct structures can be observed at positions where there are several known $N^*$ and $\Delta$ resonances with $N\pi\pi$ decay modes. Due to many ambiguities, it is not possible to assign resonances to these structures without a full partial-wave analysis of the data. For higher masses, the multitude of excited baryons creates a smooth curve which has a shoulder around $2.2\,\mathrm{GeV}/c^2$.

Essential for the partial-wave analysis are resonances in the $p\pi^\pm$ and $\pi^+\pi^-$ subsystems which appear as intermediate states, the so-called isobars. The $\pi^+\pi^-$ invariant mass distribution in Fig.~\ref{fig:mrho} shows clear signatures of $\rho^0$(770), $f_0$(980) and $f_2$(1270). A similar set of resonances was observed in the diffractive dissociation of pions into $\pi^-\pi^+\pi^-$~\cite{com10}.

The invariant mass spectrum of the $p_f\pi^-$ subsystem, depicted in Fig.~\ref{fig:mdeltan}, exhibits a distinct excited baryon spectrum, featuring a prominent $\Delta^0$(1232)$P_{33}$ together with additional structures that are probably related to the $N$(1440)$P_{11}$, $N$(1650)$S_{11}$ and $\Delta$(1700)$D_{33}$. However, also here assignments based on the mass alone are ambiguous. Naturally, the doubly charged $p_f\pi^+$ combination is less populated. In addition to the outstanding $\Delta^{++}$(1232)$P_{33}$, there seem to be higher excitations around $1.9\,\mathrm{GeV}/c^2$.

In Fig.~\ref{fig:dal1},~\ref{fig:dal2} and \ref{fig:dal3}, the three-body invariant mass is presented versus the invariant masses of the three possible sub-systems in order to illustrate the quality of the data sample. The above described isobars appear clearly as vertical bands in this manner.

\begin{figure}[ht]
  \begin{minipage}[]{.3\textwidth}
    \centering
    \includegraphics[clip,trim= 5 0 30 30, width=\textwidth]{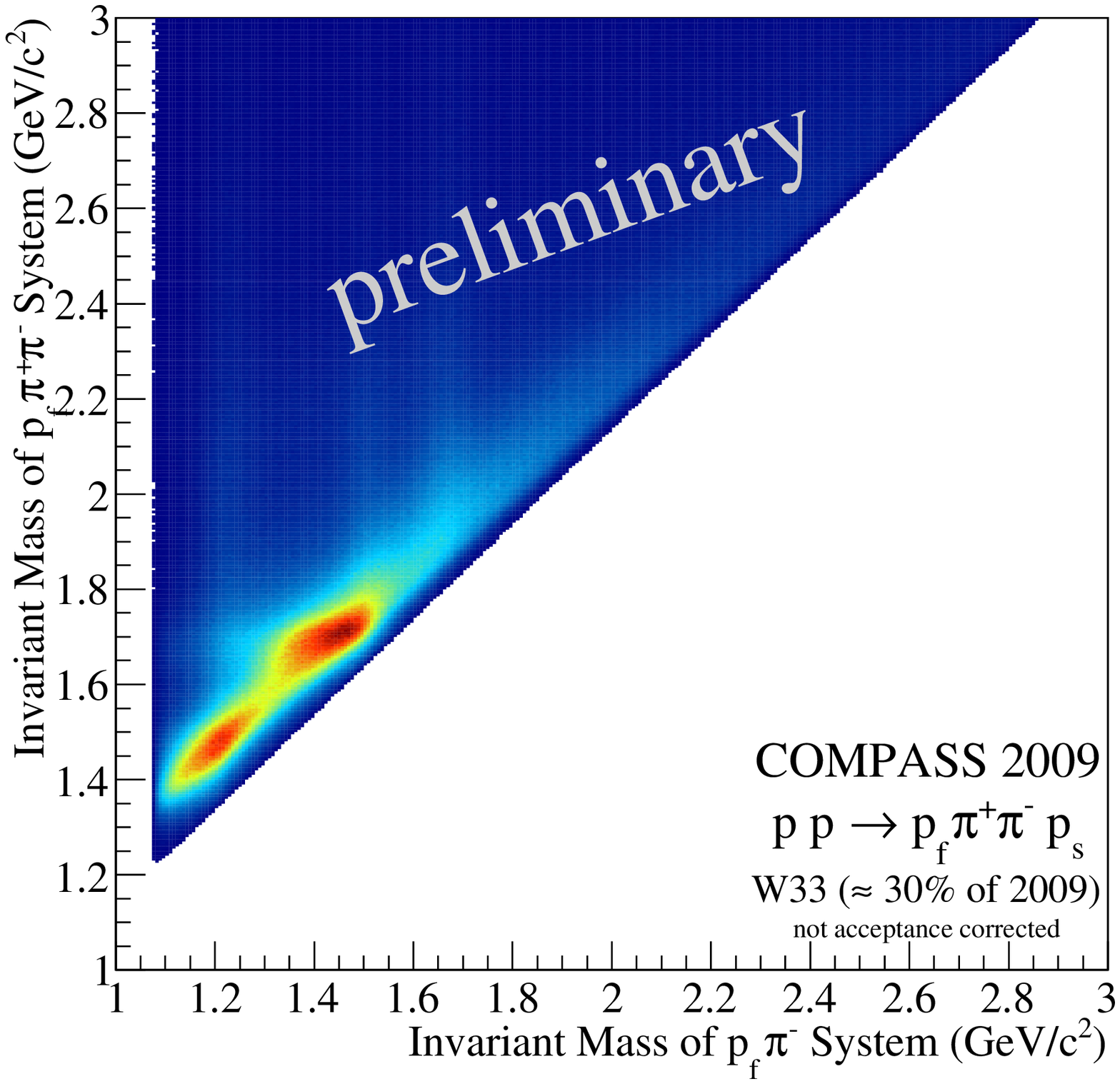}
    \caption{\em Invariant mass of $p_f\pi^+\pi^-$ vs. $p_f\pi^-$ subsystem}
    \label{fig:dal1}
  \end{minipage}
  \hfill
  \begin{minipage}[]{.3\textwidth}
    \centering
    \includegraphics[clip,trim= 5 0 30 30, width=\textwidth]{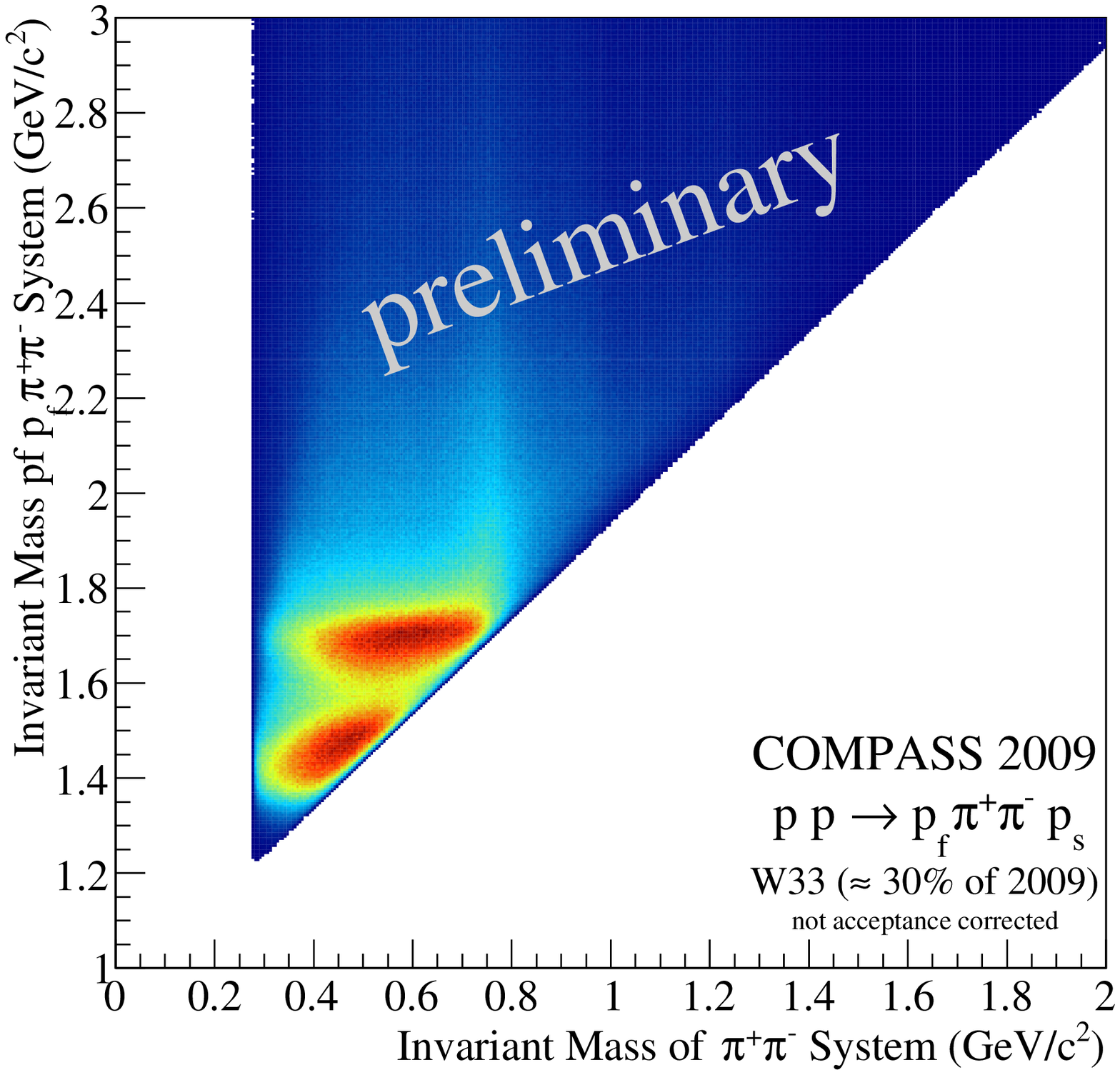}
    \caption{\em Invariant mass of $p_f\pi^+\pi^-$ vs. $\pi^+\pi^-$ subsystem}
    \label{fig:dal2}
  \end{minipage}
  \hfill
  \begin{minipage}[]{.3\textwidth}
    \centering
    \includegraphics[clip,trim= 5 0 30 30, width=\textwidth]{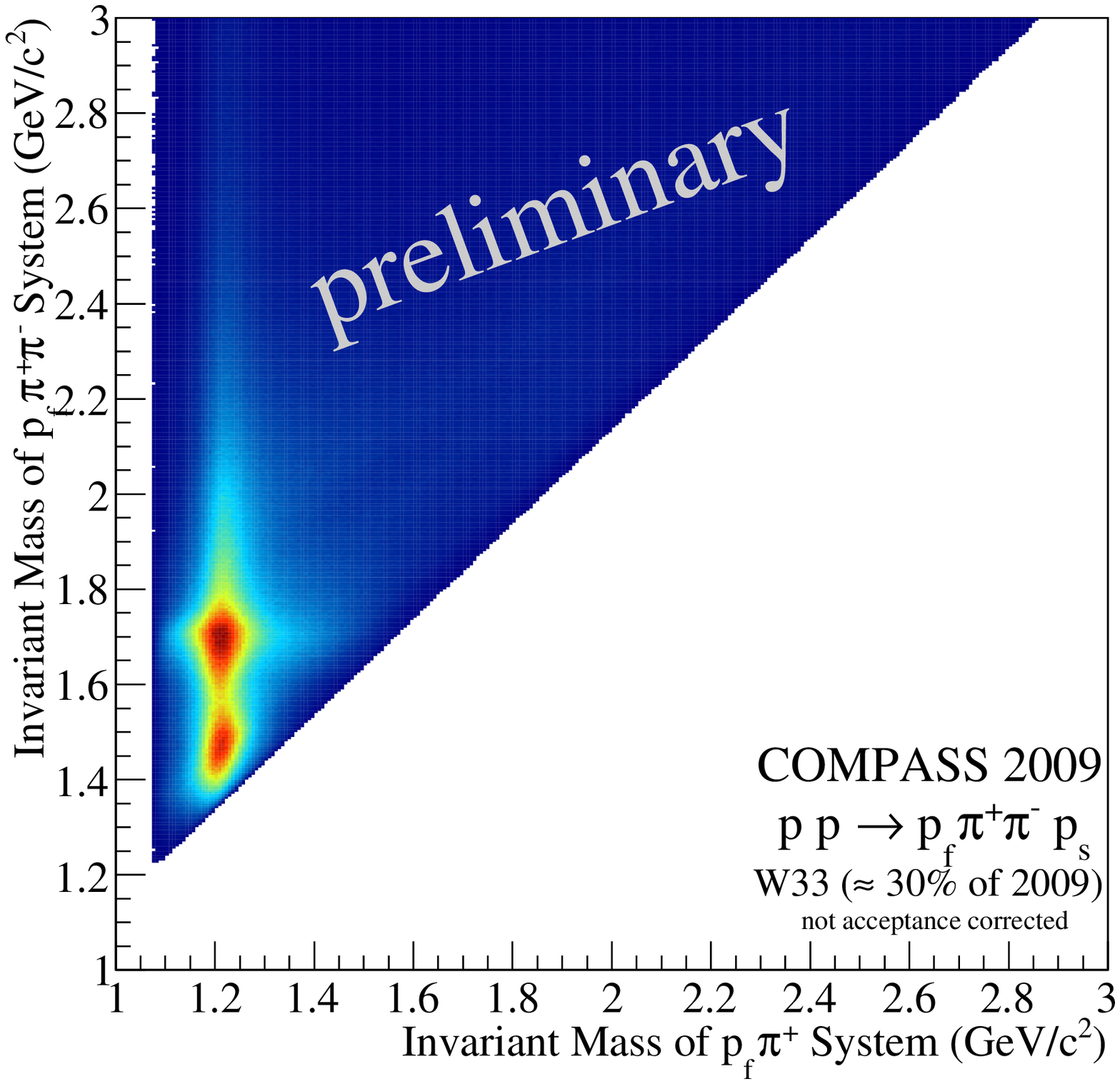}
    \caption{\em Invariant mass of $p_f\pi^+\pi^-$ vs. $p_f\pi^+$ subsystem}
    \label{fig:dal3}
  \end{minipage}
\end{figure}

\mathversion{bold}
\section{Diffractive dissociation of protons into $p_f\,K^+K^-$ final states}
\mathversion{normal}
\label{sec:kaon}

A different aspect of the baryon spectrum becomes accessible when the pions are replaced by kaons in the event selection described above. However, the number of events is considerably lower and therefore the unambiguous identification of resonances is more difficult.

\begin{figure}[t]
  \begin{minipage}[]{.5\textwidth}
    \centering
    \includegraphics[clip,trim= 10 0 40 10, width=.65\textwidth]{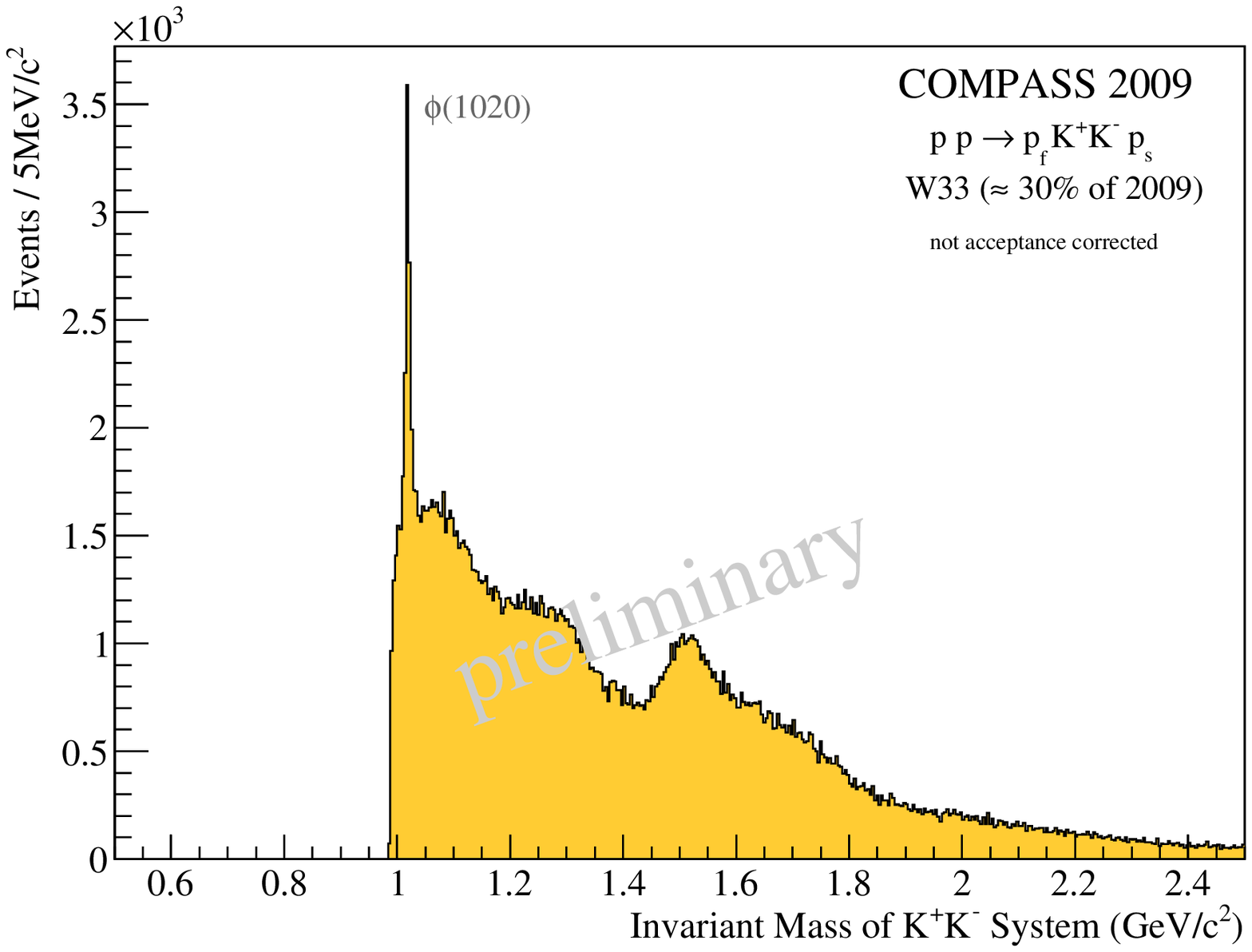}
    \caption{\em Invariant mass distribution\\ of $K^+K^-$ subsystem}
    \label{fig:mKK}
  \end{minipage}
  \hfill
  \begin{minipage}[]{.5\textwidth}
    \centering
    \includegraphics[clip,trim= 10 0 40 10, width=.65\textwidth]{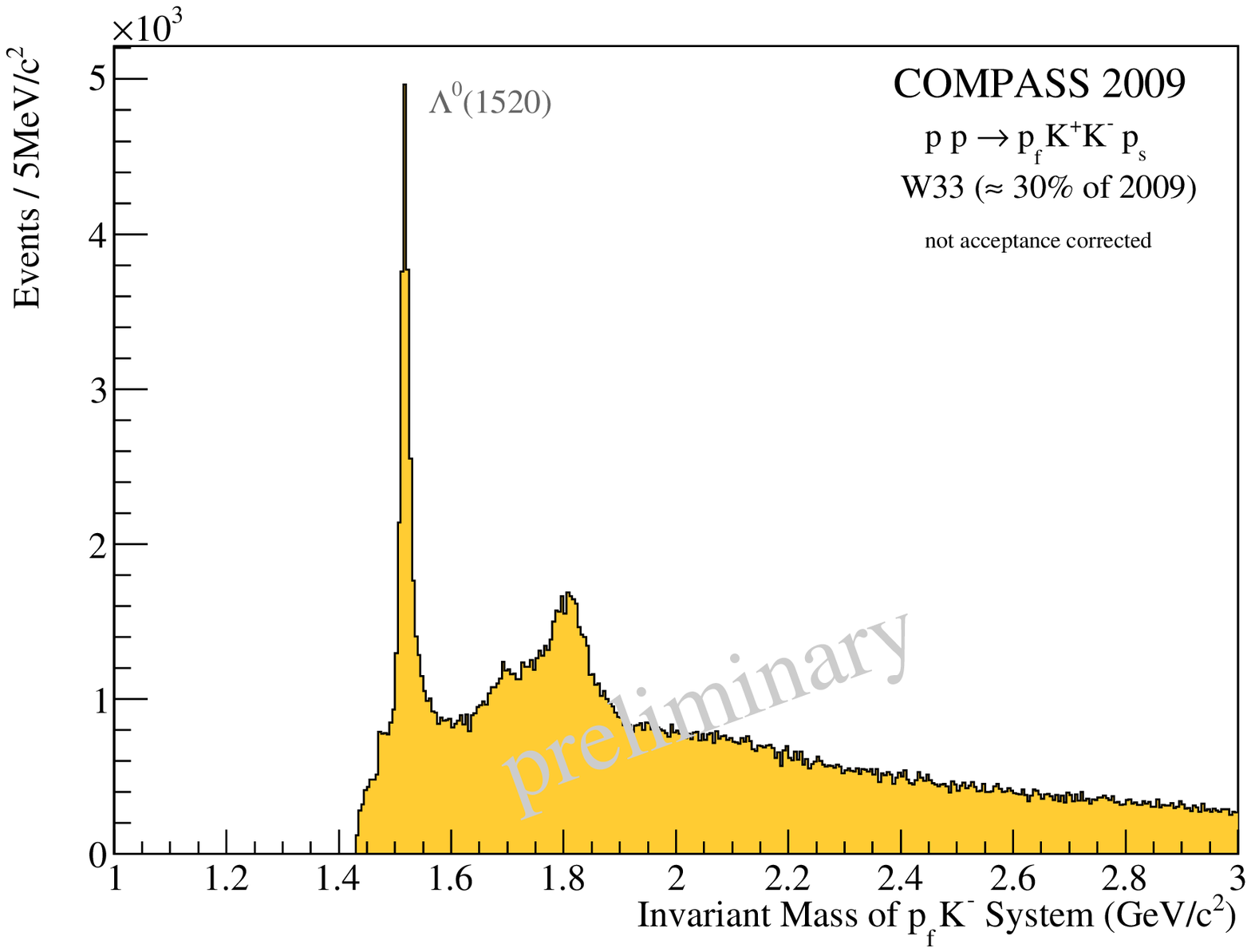}
    \caption{\em Invariant mass distribution\\ of $p_fK^-$ subsystem}
    \label{fig:mpKm}
  \end{minipage}
\end{figure}

While no special features can be seen in the three-particle invariant mass spectrum, the subsystems do show interesting structures. Most prominent is the very narrow $\phi$(1020) peak that appears as expected in the $K^+K^-$ invariant mass as shown in Fig.~\ref{fig:mKK}. In addition the invariant mass distribution exhibits structures at masses of known resonances like the $a_2$(1320) and the $f_0$(1500).

A sharp baryon resonance, the $\Lambda$(1520)$D_{03}$, can be found in the invariant mass spectrum of the $p\,K^-$ combination (cf.~Fig.~\ref{fig:mpKm}). Higher baryon excitations with strangeness are visible for example around $1.7$ and $1.8\,\mathrm{GeV}/c$, although less pronounced.


\section{Partial-Wave Analysis}
\label{sec:pwa}

The selected data set is the starting point for a dedicated partial-wave analysis. The incoming beam proton scattering off the target is excited into an intermediate state $X$, with quantum numbers which can differ from those of the initial state. This reaction can be assumed to proceed via $t$-channel Reggeon exchange, thus justifying the factorisation of the total cross section into a resonance and a recoil vertex without final state interaction. Considering only subsequent two-body decays of $X$ (i.e.~applying the isobar~model)~\cite{com10}, three different decay topologies into the same final state $p_f\,\pi^+\pi^-$ are possible which are shown in Fig.~\ref{fig:feyn}.
\begin{figure}[b]
  \begin{center}
    \includegraphics[width=.30\textwidth]{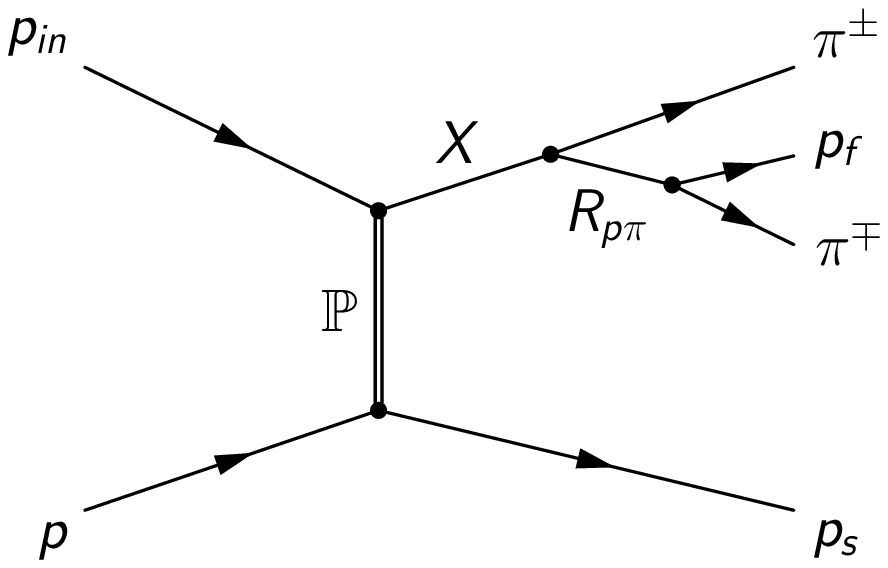}
  \hspace{1cm}
  \includegraphics[width=.30\textwidth]{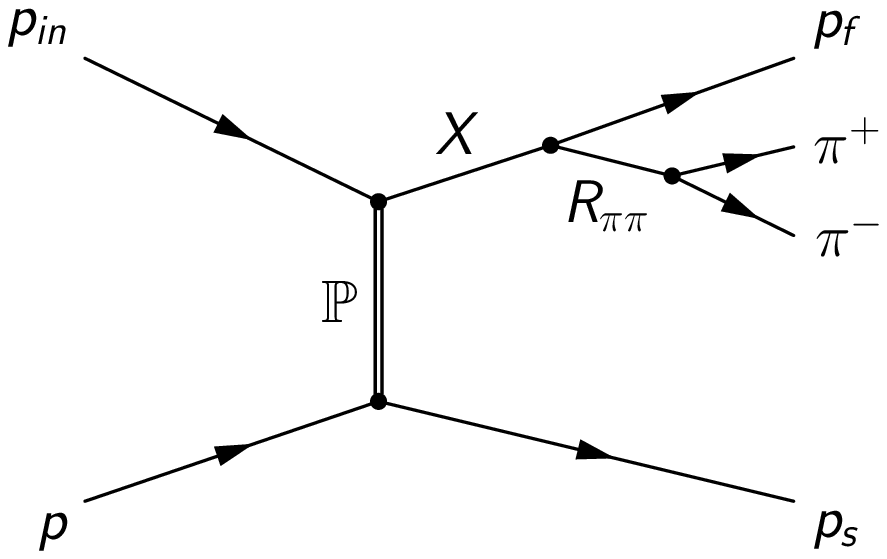}
  \end{center}
  \caption{\em Possible Decays of Resonance $X$}
  \label{fig:feyn}
\end{figure}

Taking the observed invariant mass spectra into account (cf.~Section \ref{sec:pion}), the isobar candidates can be either mesonic (e.g. $(\pi\pi)_S$, $\rho^0(770)$, $f_2(1270)$, \ldots) or baryonic (e.g. $\Delta(1232)P_{33}$, $N(1440)P_{11}$, $N(1650)S_{11}$, $\Delta(1700)D_{33}$, \ldots). The partial-wave analysis is carried out by a program developed at Brookhaven~\cite{bnl} and adapted for COMPASS~\cite{rootpwa}. $D$-functions and the canonical basis are used to evaluate the decay amplitudes.

\section{Conclusions}

In the years 2008 and 2009, the COMPASS experiment collected  a unique data set with a proton beam impinging on a liquid hydrogen target. As the diffractive dissociation of the beam proton plays a dominant role, the high resolution spectrometer combined with the clean trigger makes COMPASS an ideal tool to explore the baryon spectrum.

Thorough event selection studies led to a clean exclusive data sample, where structures at positions of known resonances become already apparent in the invariant mass distributions. Profiting from partial-wave analysis techniques developed for the search of exotic mesons~\cite{com10}, COMPASS has great potential to contribute to the field of light-quark baryon spectroscopy.

\acknowledgements{%
This work is supported by the German Bundesministerium f\"ur Bildung und Forschung, the Maier-Leibnitz-Laboratorium der LMU und TU M\"unchen, and the DFG Cluster of Excellence \textit{Origin and Structure of the Universe}.
}


%

}  


\end{document}